\documentstyle[12pt]{article}
\newcommand {\beq}{\begin{equation}}
\newcommand {\eeq}{\end{equation}}
\newcommand {\beqa}{\begin{eqnarray}}
\newcommand {\eeqa}{\end{eqnarray}}
\newcommand {\n}{\nonumber \\}

\begin{document}
\setlength{\oddsidemargin}{0cm}
\setlength{\baselineskip}{7mm}

\begin{titlepage}
 \renewcommand{\thefootnote}{\fnsymbol{footnote}}
$\mbox{ }$
\begin{flushright}
\begin{tabular}{l}
KEK-TH-833\\
Jul. 2002
\end{tabular}
\end{flushright}

~~\\
~~\\
~~\\

\vspace*{0cm}
    \begin{Large}
       \vspace{2cm}
       \begin{center}
         {Matrix Models in Homogeneous Spaces}      \\
       \end{center}
    \end{Large}

  \vspace{1cm}

\begin{center}
           Yoshihisa K{\sc itazawa}\footnote
           {
e-mail address : kitazawa@post.kek.jp}

         {\it Institute of Particle and Nuclear Studies,\\
         High Energy Accelerator Research Organization (KEK),}\\
               {\it Tsukuba, Ibaraki 305-0801, Japan} \\
\end{center}

\vfill

\begin{abstract}
\noindent
We investigate non-commutative gauge theories in homogeneous spaces
$G/H$.
We construct such theories by adding cubic terms
to IIB matrix model which
contain the structure constants of $G$.
The isometry of a homogeneous space,
$G$ must be a subgroup of $SO(10)$ in our construction.
We investigate $CP^2=SU(3)/U(2)$ case in detail
which gives rise to 4 dimensional non-commutative gauge theory.
We show that non-commutative gauge theory on $R^4$ can be realized
in the large $N$ limit by letting the action approach
IIB matrix model in a definite way.
We discuss possible relevances of these theories
to the large $N$ limit of IIB matrix model.
\end{abstract}
\vfill
\end{titlepage}
\vfil\eject

\section{Introduction}
\setcounter{equation}{0}
Matrix models which are dimensional reductions of 10 dimensional super
Yang-Mills theory first appeared as low energy effective theories of
$N$ coincident D-branes.
They may also be regarded as matrix regularization of
light-cone membrane action\cite{BFSS}
or Green-Schwarz superstring action\cite{IKKT}.
Our hope is that they may provide us second quantized string theory
in the large $N$ limit. It is because they may contain arbitrary numbers
of strings represented as block-diagonal matrices.

Non-commutative D-branes such as the following
are formal classical solutions of matrix models in the
large $N$ limit
\beqa
&&A^{cl}_1=\hat{p},~~A^{cl}_2=\hat{q},\n
&&[\hat{p},\hat{q}]=-i ,
\eeqa
since they solve the equation of motion
\beq
[A_{\mu},[A_{\mu},A_{\nu}]]=0 .
\eeq
Non-commutative (NC) gauge theory is obtained from matrix models
around NC space-time\cite{CDS}\cite{AIIKKT}\cite{Li}.
The advantage of matrix model construction of NC gauge theory
is that it maintains the manifest gauge invariance under
$U(N)$ transformations
\beq
A_{\mu} \rightarrow UA_{\mu}U^{\dagger} .
\eeq
The gauge invariant observables of NC gauge theory,
the Wilson lines were constructed through matrix
models\cite{IIKK}\cite{Gross}.
NC gauge theory exhibits UV-IR mixing which is
a characteristic feature of string theory\cite{MRS}.

In this paper we investigate NC gauge theory on homogeneous
spaces through matrix models.
It is an interesting problem on its own to
study NC gauge theories on curved manifolds.
A homogeneous space is realized as $G/H$
where $G$ is a Lie group and $H$ is a closed subgroup of $G$.
We further assume that they are symmetric spaces which
are invariant under space reflection (parity).
At Lie algebra level in symmetric spaces,
the following commutation relations hold
\beqa
&&g=h+m,\n
&&[h,h] \subset h, ~~[h,m] \subset m,
~~[m,m] \subset h,
\eeqa
where $g$ and $h$ are the generators of $G$ and $H$
respectively.

We construct NC gauge theories on homogeneous
spaces by adding cubic
couplings to IIB matrix model with a large but finite $N$.
The isometry of a homegeneous space $G/H$ is
the group $G$.
The group $G$ has to be a subgroup of $SO(10)$
which is the symmetry of IIB matrix model.
We investigate $CP^2=SU(3)/U(2)$ case in detail
which gives rise to 4 dimensional NC gauge theory.
We show that NC gauge theory on $R^4$ can be realized
in the large $N$ limit by letting the action approach
IIB matrix model in a definite way.
Although SUSY is broken in general with finite $N$,
we argue that SUSY is locally resurrected in such a limit.
We hope that our construction is useful to investigate 4 dimensional
NC super Yang-Mills theory nonperturbatively.
Since the strength of the cubic couplings formally
vanish in the large $N$ limit,
our results may be relevant to investigate the large $N$ limit of
IIB matrix model.

In section 2, we construct fuzzy homogeneous spaces $G/H$
as the orbit of a state in a definite representation
of dimension $N$ which
is invariant under $H$ modulo the overall $U(1)$ phase.
We then construct NC gauge fields as bi-local states.
Such a construction facilitates matrix model realizations
since gauge fields  are naturally represented by $N\times N$  Hermitian
matrices. In sections 3, we construct NC gauge theories
on homogeneous spaces $G/H$ through matrix models.
For this purpose we deform
IIB matrix model by adding cubic terms in $A_{\mu}$
which contain structure constants of $G$.
We investigate $CP^2=SU(3)/U(2)$ case in detail
which gives rise to 4 dimensional NC gauge theory.
We show that NC gauge theories on $R^4$ can be realized
in the large $N$ limit by letting the action approach
IIB matrix model in a definite way.
We investigate quantum effects of 4 dimensional NC gauge theory
and its formal infrared limit.
We conclude in section 4 with discussions.
We discuss possible relevances of these theories
to the large $N$ limit of IIB matrix model.

\section{Non-commutative Spacetime}
\setcounter{equation}{0}

In this section, we formulate a generic
procedure to construct fuzzy homogeneous spaces $G/H$
and gauge fields on them.
We pick a state $|0>$ in a definite representation of
$G$ which is invariant under $H$.
In this construction, we identify the states which only differ
by their $U(1)$ phases since they are equivalent as quantum states.
The set of all states which can be reached by multiplying
elements of $G$ to $|0>$ is called the orbit of $|0>$.
Fuzzy homogeneous spaces $G/H$ are constructed
as the orbit of $|0>$.
It is represented by the irreducible representation
which is descended from $|0>$.
The dimension of the representation $N$ can be interpreted as the
volume of fuzzy $G/H$ in the unit of non-commutativity scale.
We recall that the basic degrees of freedom
in NC gauge theory are bi-local fields
which can be interpreted as zeromodes of open strings\cite{IKK}.
We then construct NC gauge fields by forming
the tensor product of the relevant irreducible representation and its
complex conjugate.
Such a construction facilitates matrix model realizations
since gauge fields are naturally represented by $N\times N$  Hermitian
matrices.

We first recall a fuzzy flat manifold $R^2$.
We have non-commutative coordinates $\hat{x},\hat{y}$ which
satisfy the canonical commutation relation
\beq
[\hat{x},\hat{y}]=i .
\eeq
Obviously we cannot realize such a commutation relation
with finite size matrices.
\footnote{
This commutation relation is realized by the guiding center
coordinates of electrons in magnetic field.
In fact NC gauge theory may be realized in quantum
Hall system\cite{Suss}.}
We can construct the creation and annihilation
operators $\hat{a},\hat{a}^{\dagger}$ out of them:
\beqa
&&\hat{a}={1\over \sqrt{2}}(\hat{x}+i\hat{y}),
~~\hat{a}^{\dagger}={1\over \sqrt{2}}(\hat{x}-i\hat{y}),\n
&&[\hat{a},\hat{a}^{\dagger}]=1 .
\label{fltcom}
\eeqa
A localized state $|0>$ at the origin can be defined as
\beq
\hat{a}|0>=0 .
\eeq
We can construct surrounding states by applying the creation
operators to $|0>$ as
\beq
\hat{a}^{\dagger}|0>,(\hat{a}^{\dagger})^2|0>,
\cdots .
\eeq
The states which can be constructed from $|0>$ by
applying $\hat{a}^{\dagger}$ finite times
are localized around the origin.
They are uniformly distributed
with the density $1/2\pi$ in accordance with the
semiclassical quantization condition.

The conjugate momentum operators are identified with
\beq
\hat{p}_{x}=\hat{y},
~~\hat{p}_{y}=-\hat{x},
\eeq
since they satisfy the canonical commutation relationship
with $\hat{x},\hat{y}$. They generate the translations
along fuzzy plane.
We subsequently introduce the adjoint operators
$P_{\mu}=[\hat{p}_{\mu},~]$. Since $P_x$ and $P_y$ commute,
we can simultaneously diagonalize them.
The eigen states can be constructed as
\beqa
P_{\mu}exp(ik_x\hat{x}+ik_y\hat{y})
&\equiv &[\hat{p}_{\mu},exp(ik_x\hat{x}+ik_y\hat{y})]\n
&=&k_{\mu}exp(ik_x\hat{x}+ik_y\hat{y}) .
\eeqa
They are the analogous of the plane waves in non-commutative space.
They can also be interpreted as the bi-local fields
or dipoles with the length $|k|$.

We next review $S^2=SU(2)/U(1)$ case as the first example of
a curved manifold.
In $SU(2)$, we have the Hermitian operators
$\hat{j}_x,\hat{j}_y,\hat{j}_z$ which satisfy the commutation relations
of the angular momentum.
\beq
[\hat{j}_x,\hat{j}_y]=i\hat{j}_z .
\eeq
Contrary to $R^2$ case, such a commutation relation
can be realized with finite size matrices
since $S^2$ is compact.
The raising and lowering operators $\hat{j}^+,\hat{j}^-$ can be formed from
$\hat{j}_x,\hat{j}_y$ which satisfy
\beqa
&&\hat{j}^{\pm}={1\over \sqrt{2}}(\hat{j}_x\pm i\hat{j}_y),\n
&&[\hat{j}^+,\hat{j}^-]=\hat{j}_z ,\n
&&[\hat{j}_z, \hat{j}^{\pm}]=\pm \hat{j}^{\pm} .
\label{angcom}
\eeqa
We consider the $N=2l+1$ dimensional
representation of spin $l$.
Since we can diagonalize $\hat{j}_z$, we can label the states
with the eigenvalue of $\hat{j}_z$ as $|m>$
where
\beq
\hat{j}_z|m>=m|m> .
\eeq
We consider the semiclassical limit where $l$ is assumed to be
large. In such a situation we expect to recover a large
smooth sphere. We can further expect that a flat fuzzy
plane is realized in the neighborhood of a particular point
such as the north-pole. In fact
the notion of locality can be introduced
in the large $l$ limit as follows.

As a localized states at the north-pole, we can choose $|l>$.
We can subsequently construct other states by using the lowering operators
\beq
\hat{j}^-|l>,(\hat{j}^-)^2|l>,\cdots .
\eeq
Since we have assumed that $l$ is large, these states
which are constructed from $|l>$ with finite
operations of $\hat{j}^-$ are all localized
around the north-pole.

For these states,
we can approximate $\hat{j}_z \sim l$.
By rescaling the operators, we can obtain the following
commutation relations from (\ref{angcom}),
\beqa
&&\tilde{a}={1\over \sqrt{l}}\hat{j}^+,
~\tilde{a}^{\dagger}={1\over \sqrt{l}}\hat{j}^-,
~\tilde{1}={1\over l}\hat{j}_z\n
&&[\tilde{a},\tilde{a}^{\dagger}] = \tilde{1},
~~[\tilde{1},\tilde{a}]={1\over l}\tilde{a},
~~[\tilde{1},\tilde{a}^{\dagger}]=-{1\over l}\tilde{a}^{\dagger} .
\label{s2com}
\eeqa
Since we obtain the identical algebra with  (\ref{fltcom}),
we conclude that flat fuzzy plane is realized around the
north-pole of $S^2$ in the large $l$ limit.
The density of the states is $1/2\pi l$ in the original
coordinate and the area of $S^2$ is $4\pi l^2$.
Thus the total number of the states is $2l$ semiclassically which is
consistent with the spin $l$ representation of $SU(2)$.

The local coordinates which satisfy the canonical commutation
relationship can be identified as
\beq
\tilde{x}={1\over \sqrt{l}}\hat{j}_x,
~~\tilde{y}={1\over \sqrt{l}}\hat{j}_y .
\eeq
The conjugate momentum operators are
\beq
\tilde{p}_x={1\over \sqrt{l}}\hat{j}_y,
~~\tilde{p}_y=-{1\over \sqrt{l}}\hat{j}_x .
\eeq
In the local patch, the eigen functions of the adjoint
$\tilde{P}_{\alpha}=[\tilde{p}_{\alpha},~]$ can be constructed just like the
flat space case as $exp(ik_x\tilde{x}+ik_y\tilde{y})$.
We point out here that adjoint operators of $J_{\mu}=[\hat{j}_{\mu},~]$
also satisfy the commutation relation of $SU(2)$.
Using the commutation relations in (\ref{s2com}) which are realized around
the north-pole, we find that $\tilde{P}^2$ can be related to the Casimir
operator of
$SU(2)$ in a local patch as follows
\beq
{J^2\over l} exp(ik_x\tilde{x}+ik_y\tilde{y})
=\tilde{P}^2 exp(ik_x\tilde{x}+ik_y\tilde{y}) +O({1\over l}) .
\eeq
The homogeneous space $S^2$ is generated by
$\hat{j}_x$ and $\hat{j}_y$ starting from the state $|l>$.
The state $|l>$ only changes its $U(1)$ phase under the rotation
around the $z$ axis.
As quantum states, we may identify the states which differ only by
their $U(1)$ phases.
Therefore fuzzy $S^2$ can also be realized as $G/H$ where $G=SU(2)$ and
$H=U(1)$.

We can parameterize $CP^1$ or $S^2$ by two complex coordinates
$u_{\alpha}$ such that
\beq
u^*_{\alpha}u_{\alpha}=1 .
\eeq
Fuzzy $S^2$ can be represented by the following
states with spin $l$.
\beq
u_{\alpha_1}\ldots u_{\alpha_{2l}} .
\eeq
We construct gauge fields on fuzzy $S^2$ as bi-local fields.
Hence the bi-local fields are represented as
\beq
u_{\alpha_1}\ldots u_{\alpha_{2l}}u^*_{\beta_1}\ldots u^*_{\beta_{2l}} .
\eeq
They can be decomposed into the irreducible representations
with spin $n$
\beq
\sum_{n=0}^{2l}(u_{\alpha_1}\ldots u_{\alpha_{n}}u^*_{\beta_1}\ldots
u^*_{\beta_{n}})
\eeq
which are traceless under the contractions of $\alpha$ and $\beta$
indices.

The adjoint generators of $SU(2)$ transformations are represented as
\beq
J^i=-{1\over 2}\left[u_{\beta}\sigma^i_{\alpha\beta}
{\partial\over\partial u_{\alpha}}
-u^*_{\alpha}\sigma^i_{\alpha\beta}
{\partial\over\partial u^*_{\beta}}\right]
\eeq
which satisfy the $SU(2)$ algebra
\beq
[J_i,J_j]=i\epsilon_{ijk}J_k .
\eeq
The Casimir operator acts on the bi-local fields as
\beq
J^2={1\over 2}(u_{\alpha}{\partial\over \partial u_{\alpha}})^2
+{1\over 2}u_{\alpha}{\partial\over \partial u_{\alpha}}
+c.c.
=n^2+n .
\eeq
Therefore the eigenvalues of the Laplacian on $S^2$, ${J^2/ l}$ are
$n(n+1)/l$ with
$n$ integers.
Since $\sum_{n=0}^{n=2l}(2n+1)=(2l+1)^2$,
a group of representations with spins up to $2l$ form the
complete basis of $N\times N$ Hermitian matrices
as it is evident from our construction.

After reviewing the well-known example of fuzzy $S^2$ case,
we can formulate a general procedure to construct
fuzzy homogeneous space $G/H$ as follows.
We first consider a representation of $G$ which contains
a (highest weight) state which is invariant under $H$ modulo the over all
$U(1)$  phase.
We also require that fuzzy $R^n$ where $n$ is the dimension of $G/H$
is realized in the local patch around such a state.
We are thus restricted to symplectic manifolds. It is because
the $\star$ product on such a manifold can be
reduced to that of a flat manifold (Moyal product) locally by choosing
the Darboux coordinates.
Since the K\"{a}hler form serves as the symplectic form,
K\"{a}hler manifolds such as $CP^n$ satisfy this requirement\cite{Masuda}.

We embed the Lie generators of $G$ into
$N$ dimensional Hermitian matrices where $N$ is the
dimension of the representation.
The gauge fields on fuzzy $G/H$ are constructed as bi-local fields.
The bi-local fields are the tensor products
of the relevant representation and the complex conjugate of it.
Since they are reducible, we can decompose them into the irreducible
representations.
They are guaranteed to form the complete basis of $N\times N$
Hermitian matrices by construction.
In what follows we discuss several concrete examples
in higher dimensions
following this general procedure.

Our main interest in this paper is $CP^2=SU(3)/U(2)$\cite{KN}.
$SU(3)$ is generated by
$8$ Hermitian operators $t^a$ which satisfy
\beq
[t^a,t^b]=if^{abc}t^c .
\eeq
The structure constants of $SU(3)$ are
\beqa
&&f_{123}=1,\n
&&f_{147}=f_{246}=f_{257}=f_{345}=f_{516}=f_{637}={1\over 2},\n
&&f_{458}=f_{678}={\sqrt{3}\over 2} .
\eeqa
The irreducible  representations of $SU(3)$ can be classified
by the Young Tableaux which is specified with a pair of integers
$(p,q)$.\footnote{There are $p+q$ boxes in the first low
and $q$ boxes in the second low.}

We consider the representation $(p,0)$ which can be realized
by totally symmetrizing the following states
\beq
u_{\alpha_1}u_{\alpha_2}\cdots u_{\alpha_p},
\eeq
where $\sum_{\alpha =1}^3u^*_\alpha u_\alpha =1$.
Let us consider the state $u_{\alpha_i}=\delta_{\alpha_i,3}$ for all $i$.
Such a state $|p>$ is a singlet of $SU(2)$ subgroup of $SU(3)$
which are generated by $t^1,t^2,t^3$ and the eigen state of
$t^8$ with the eigenvalue ${p /\sqrt{3}}$.

The commutation relations among $t^4\cdots t^7$ are
\beqa
&&[t^4,t^5] = i{\sqrt{3}\over 2} t^8 +i{1\over 2}t^3,\n
&&[t^6,t^7] = i{\sqrt{3}\over 2} t^8 -i{1\over 2}t^3.
\eeqa
We may introduce the raising and lowering operators:
\beq
u^{\pm}={1\over \sqrt{2}}(t^6\pm it^7),
~~v^{\pm}={1\over \sqrt{2}}(t^4\pm it^5).
\eeq
As $S^2$ case, we can construct descendent states
by applying lowering operators $u^{-},v^{-}$ to $|p>$.
\beq
|p>,u^{-}|p>,v^{-}|p>,(u^{-})^2|p>,(v^{-})^2|p>,u^{-}v^{-}|p>,\cdots .
\eeq
In the large $p$ limit, the states which are obtained
from $|p>$ with finite actions of the lowering operators
form a localized patch.

After rescaling $u,v\rightarrow \sqrt{p/2}\tilde{u},\sqrt{p/2}\tilde{v}$
we obtain the following commutation relation which are realized
in such a local patch,
\beqa
&&[\tilde{u}^+,\tilde{u}^-] = 1 +O(1\sqrt{p}),\n
&&[\tilde{v}^+,\tilde{v}^-] = 1 +O(1\sqrt{p}).
\label{cp2com}
\eeqa
We thus conclude that flat fuzzy $R^4$ is realized
in a local patch of $CP^2$ in the large $p$ limit.
The density of the states is $1/\pi^2 p^2$ in the original
coordinate and the volume of $CP^2$ is $\pi^2 p^4/2$.
Thus the total number of the states is $p^2/2$ semiclassically which is
consistent with the $(p,0)$ representation of  $SU(3)$.

The local coordinates which satisfy the canonical commutation
relationship can be identified as
\beqa
&&\tilde{x}={\sqrt{2}\over \sqrt{p}}t^4,
~~\tilde{y}={\sqrt{2}\over \sqrt{p}}t^5,
~~[\tilde{x},\tilde{y}]=i,\n
&&\tilde{w}={\sqrt{2}\over \sqrt{p}}t^6,
~~\tilde{z}={\sqrt{2}\over \sqrt{p}}t^7,
~~[\tilde{w},\tilde{z}]=i.
\label{cp2com}
\eeqa
We also find
\beqa
&&[t^1,\tilde{x}]={i\over 2}\tilde{z},
~~[t^1,\tilde{y}]=-{i\over 2}\tilde{w},
~~[t^1,\tilde{w}]={i\over 2}\tilde{y},
~~[t^1,\tilde{z}]=-{i\over 2}\tilde{x},\n
&&
[t^2,\tilde{x}]={i\over 2}\tilde{w},
~~[t^2,\tilde{y}]={i\over 2}\tilde{z},
~~[t^2,\tilde{w}]=-{i\over 2}\tilde{x},
~~[t^2,\tilde{z}]=-{i\over 2}\tilde{y},\n
&&
[t^3,\tilde{x}]={i\over 2}\tilde{y},
~~[t^3,\tilde{y}]=-{i\over 2}\tilde{x},
~~[t^3,\tilde{w}]=-{i\over 2}\tilde{z},
~~[t^3,\tilde{z}]={i\over 2}\tilde{w}.
\eeqa
We may thus identify the $SU(2)$ subgroup
formed by $t^1,t^2,t^3$ as
a subgroup of $SO(4)$ as
\beq
t^1={-i\over 2}(L^{14}-L^{23}),
~~t^2={-i\over 2}(L^{13}+L^{24}),
~~t^3={-i\over 2}(L^{12}-L^{34}) .
\eeq

The conjugate momentum operators are
\beqa
&&\tilde{p}_x=\tilde{y},
~~\tilde{p}_y=-\tilde{x},\n
&&\tilde{p}_w=\tilde{z},
~~\tilde{p}_z=-\tilde{w} .
\label{4dmom}
\eeqa
In the flat space limit,
the eigen functions of the adjoint operators
$\tilde{P}_{\alpha}=[\tilde{p}_{\alpha},~]$ can be constructed
as $exp(ik\cdot x)\equiv
exp(ik_x\tilde{x}+ik_y\tilde{y}+ik_w\tilde{w}+ik_z\tilde{z})$.
We introduce
here the adjoint operators $T^a=[t^a,~]$ which also satisfy the
commutation relation of $SU(3)$.
Using the commutation relations in (\ref{cp2com}) which are realized around
the north-pole, we find that $\tilde{P}_{\alpha}^2$ can be related to the
Casimir operator of
$SU(3)$ in a local patch
\beq
{2\over p}(T^a)^2 exp(ik\cdot x)
=\tilde{P}_{\alpha}^2 exp(ik\cdot x) +O({1\over p}) .
\eeq
Thus the Laplacian on $CP^2$ which reduces to that of flat
$R^4$ in a local patch in the large $N$ limit is $2(T^a)^2/p$.
The eigenvalues $2(T^a)^2/p$ are $2n(n+2)/p$
for the representation $(n,n)$.

It is because
the bi-local states are represented as
\beq
u_{\alpha_1}\ldots u_{\alpha_{p}}u^*_{\beta_1}\ldots u^*_{\beta_{p}}.
\eeq
They can be decomposed into the irreducible representations
of $(n,n)$ type
\beq
\sum_{n=0}^p (u_{\alpha_1}\ldots u_{\alpha_{n}}u^*_{\beta_1}\ldots
u^*_{\beta_{n}})
\eeq
which are traceless under the contractions of any pairs of
$\alpha$ and $\beta$ indices.
The adjoint generators of $SU(3)$ transformations are represented as
\beq
T^a=-\left[u_{\beta}t^a_{\alpha\beta}
{\partial\over\partial u_{\alpha}}
-u^*_{\alpha}t^a_{\alpha\beta}
{\partial\over\partial u^*_{\beta}}\right]
\eeq
which satisfy the $SU(3)$ algebra.
The Casimir operator acts on the bi-local fields as
\beq
T^2={1\over 2}(u_{\alpha}{\partial\over \partial u_{\alpha}})^2
+u_{\alpha}{\partial\over \partial u_{\alpha}}
+c.c.
=n^2+2n .
\eeq
The dimension of the
representation $(n,n)$ is $(n+1)^3$.
Since $\sum_{n=0}^p(n+1)^3= (p+1)^2(p+2)^2/4$,
a group of the irreducible representations $(n,n)$ with $n$ up to $p$
form a complete basis of $N\times N$ Hermitian matrices where
$N=(p+1)(p+2)/2$ is the
dimension of the representation $(p,0)$.

Our final example in this section is
$CP^3=SO(5)/U(2)$\cite{ZH}$\sim$\cite{Kimura}.
$SO(5)$ are generated by 10 anti-symmetric matrices $t_{\mu\nu}$
which satisfy:
\beq
[t_{\mu\nu},t_{\rho\sigma}]
=i\delta_{\mu\rho}t_{\nu\sigma}-i\delta_{\nu\rho}t_{\mu\sigma}
-i\delta_{\mu\sigma}t_{\nu\rho}+i\delta_{\nu\sigma}t_{\mu\rho} .
\eeq
Our investigation is constrained to the homogeneous spaces
which can be realized through IIB matrix model.
$G$ is maximal in this case since the number of its generators
cannot exceed 10 which is the number of bosonic matrices
$A_{\mu}$ in IIB matrix model.
$SO(5)$ contains a subgroup $SO(4)=SU(2)\times SU(2)$
which is generated by
\beqa
&&j_1={1\over 2}(t_{23}+t_{14}),
~~j_2={1\over 2}(t_{31}+t_{24}),
~~j_3={1\over 2}(t_{12}+t_{34}),\n
&&
k_1={1\over 2}(t_{23}-t_{14}),
~~k_2={1\over 2}(t_{31}-t_{24}),
~~k_3={1\over 2}(t_{12}-t_{34}).
\eeqa
We can simultaneously diagonalize $t^2, j^2, k^2, j_z, k_z$.

The irreducible representations of $SO(5)$ are represented
by the Young Tableaux with a pair of integers $(m,n)_5$ with $m\geq
n$. We consider the representation $(p/2,p/2)_5$ which
decomposes into the representations of $SO(4)=SU(2)\times SU(2)$
as follows
\beq
({p\over 2},{p\over 2})_5= (0,{p\over 2})_4
+({1\over 2},{p-2\over 2})_4
+\cdots +({p\over 2},0)_4 .
\eeq
As a localized state $|p>$, we consider such a state
in $({p/2},0)_4$ which is invariant under
$U(2)$ modulo $U(1)$ phases:
\beqa
&&j^2|p>=({p\over 2})({p+2\over 2})|p>, ~~j_3|p>=({p\over 2})|p>,\n
&&k^2|p>=0, \n
&&t^2|p>=({p^2+4p\over 2})|p> .
\eeqa
In a local patch around $|p>$, the following commutation relations
are realized
\beqa
&&[j_1,j_2]=ij_3\sim i{p\over 2},\n
&&[t_{15},t_{25}]=it_{12}\sim i{p\over 2},\n
&&[t_{35},t_{45}]=it_{34}\sim i{p\over 2} .
\eeqa
We thus find that fuzzy $R^6$ is realized in the
large $p$ limit in the local patch just like
the previous examples.

The bi-local fields are obtained by
considering the direct product of
\beq
(p/2,p/2) \otimes (p/2,p/2)
\eeq
which can be decomposed into the irreducible
representations of $SO(5)$ as
\beq
\sum_{m=0}^p\sum_{n=0}^m(m,n).
\eeq
Since the dimension of the $(m,n)$ representation is
\beq
D(m,n)=(1+m-n)(1+2n)(2+m+n)(3+2m)/6 ,
\eeq
we can check that
the dimension of the bi-local fields
agrees with those of $N\times N$ Hermitian matrices
\beq
\sum_{m=0}^p\sum_{n=0}^m D(m,n)=D(p/2,p/2)^2 .
\eeq

\section{Matrix Model Realization}
\setcounter{equation}{0}

In this section, we construct NC gauge theories
on homogeneous spaces $G/H$ through matrix models.
We propose to deform IIB matrix model action as follows
generalizing the $S^2=SU(2)/U(1)$ case\cite{IKTW}
\footnote{See also\cite{Nair}\cite{Bonel}.}:
\beq
S_{IIB}\rightarrow S_{IIB}+
{i\over 3}\alpha f_{\mu\nu\rho}Tr[A_{\mu},A_{\nu}]A_{\rho} ,
\label{act2}
\eeq
where $f_{\mu\nu\rho}$ is the structure constant of $G$.
Since there are 10 Hermitian matrices $A_{\mu}$ in IIB matrix model,
the number of the Lie generators of $G$ cannot exceed 10 in this
construction.
This action does not preserve SUSY unless $G=SU(2)$.
However
we show that NC gauge theory on $R^4$ can be realized
in the large $N$ limit by letting the action approach
IIB matrix model in a definite way.
Although SUSY is broken in general with finite $N$,
we argue that SUSY is locally resurrected in such a limit.
Since this model possesses the translation invariance
\beq
A_{\mu}\rightarrow A_{\mu}+c_{\mu}
\eeq
and also
\beq
\psi\rightarrow \psi+\epsilon,
\eeq
we remove these zero-modes by restricting $A_{\mu}$ and $\psi$
to be traceless.

The equation of motion is
\beq
[A_{\mu},[A_{\mu},A_{\nu}]]+i\alpha f_{\mu\rho\nu}[A_{\mu},A_{\rho}]=0 .
\eeq
The nontrivial classical solution is
\beq
A_{a}^{cl}=\alpha t^{a},
~~other ~A_{\mu}^{cl} =0 ,
\label{clsol}
\eeq
where $t^{a}$'s satisfy the Lie algebra of $G$.
Although diagonal matrices also solve the equation of
motion, the nontrivial solution (\ref{clsol}) minimizes the classical action.
In $SU(2)$ case, it is evaluated
for the irreducible representation of spin $l$
\beq
-{\alpha^4\over 6}l(l+1)(2l+1) .
\eeq
For a large but fixed $N$,the irreducible representation
of spin $l$ where $N=2l+1$ is selected by minimizing
the classical action\cite{Myers}.
There is no quantum corrections to worry about
thanks to SUSY.

Let us investigate the analogous problem in $CP^2=SU(3)/U(2)$ case.
The classical action for the irreducible representation
$(p,q)$ of $SU(3)$ is evaluated as
\beq
-{\alpha^4\over 4}C_2(p,q)dim(p,q),
\label{clsact}
\eeq
where $C_2(p,q)$ is the Casimir and $dim(p,q)$ is the dimension
of the representation
\beqa
&&C_2(p,q)={1\over 3}[p(p+3)+q(q+3)+pq],\n
&&dim(p,q)={(p+1)(q+1)(p+q+2)\over 2} .
\eeqa
We can see that the classical action is
minimized for the $(p,0)$ type representation
with the fixed $dim(p,q)=N$.
We also note that reducible representations
do not minimize the classical action.
We conclude that the desired representation
of $(p,0)$ type which is relevant to fuzzy $CP^2$ is selected by minimizing
the classical action.
Therefore this model realizes $U(1)$ NC gauge theory on $CP^2$
with finite $N$.

In the large $N$ limit, the fuzzy $R^4$ is realized locally
as it has been shown in the previous section.
We expand the action around the classical solution as
$A_{\mu}=\alpha\sqrt{p/2}(\hat{p}_{\mu}+\hat{a}_{\mu})$.
In this parameterization, the non-commutativity scale
is fixed to be 1.
After using the Moyal-Weyl correspondence,
\beqa
&&\hat{a}\rightarrow a(x),\n
&&\hat{a}\hat{b}\rightarrow a(x)\star b(x),\n
&&Tr\rightarrow ({1\over 2\pi})^2\int d^4x ,
\eeqa
we obtain the following NC gauge theory from (\ref{act2})
\beqa
&&-\alpha^4( {p\over 2})^2({1\over 2\pi})^2\int d^4x
\left({1\over 4}[D_{\alpha},D_{\beta}]^2
+{1\over 2}[D_{\alpha},\phi_i]^2
+{1\over 4}[\phi_i,\phi_j]^2\right.\n
&&\left.+{1\over 2}\bar{\psi}\Gamma_{\alpha}[D_{\alpha},\psi]
+{1\over 2}\bar{\psi}\Gamma_{i}[\phi_i,\psi] \right)_{\star} .
\label{nc4act}
\eeqa
The cubic terms are suppressed by $\sqrt{2/p}$.
In this way, we identify the coupling constant of NC gauge theory
as
\beq
g^2_{NC}=({4\pi\over p\alpha^2})^2 .
\eeq
The analogous relations hold for 2 and 6 dimensional cases
respectively as
\beq
g^2_{NC_2}=2\pi ({1\over l\alpha^2})^2,
~~g^2_{NC_6}=(2\pi)^3({2\over p\alpha^2})^2 .
\label{coupls}
\eeq
The classical action assumes the following value in the large $N$ limit
\beq
-{2\pi^2p^2\over 3g_{NC}^2}\sim -{4\pi^2N\over 3g_{NC}^2} .
\eeq

We need to choose $g_{NC}\sim 1$ to obtain
interacting NC gauge theory.
For this purpose, we may choose $\alpha$
such that $g_{NC} \sim 1$ for a fixed $N$.
We therefore generally need to choose $\alpha^2 \sim 1/p$.
In 4 dimensions, we find $N\sim p^2/2$ for the $(p,0)$ representation which
minimizes the classical action.
If we let $N$ large in this way, we find that $\alpha$
vanishes in the large $N$ limit as $\alpha^2\sim O(\sqrt{1/N})$.
From (\ref{nc4act}), we can see that SUSY is locally recovered
in this limit.
We have argued that NC gauge theory on $R^4$ can be obtained
by expanding IIB matrix model around fuzzy $R^4$\cite{AIIKKT}.
Our matrix model construction makes such a statement
more precise.
Although we can formulate NC gauge theories nonperturbatively
through unitary matrices\cite{AMNS}, our construction
may be useful for supersymmetric gauge theories.

We move on to investigate quantum theory.
After the gauge fixing,
the quadratic action for $\hat{a}_{\mu}$ is simply given by
\beq
{1\over 2g_{NC}^2}
(2\pi)^2Tr\left(\hat{a}_{\nu}P_{\mu}P_{\mu}\hat{a}_{\nu}
\right) .
\eeq
As we have described in the preceding section, we may identify the
eigenstates of $P_{\mu}^2=2T^2/p$ with $(n,n)$ representations of $SU(3)$
with the eigenvalues
\beq
2n(n+2)/p .
\eeq
In a local patch, we can adopt the four dimensional approximation
for  $\hat{p}_{\mu}$ as in (\ref{4dmom}).
In such an approximation, the eigenvalues $\tilde{P}_{\alpha}^2$ are
$2n^2/p$.  They are uniformly distributed over
the momentum space with the density $p^2/8\pi^2$.

The one loop correction to the classical action is
\beq
{1\over 2}Trlog(P^2\delta_{\mu\nu})-Trlog(P^2)
-{1\over 4}Trlog\left( (P^2+{i\over 2}F_{\mu\nu}\Gamma^{\mu\nu})
({1+\Gamma_{11}\over 2})\right),
\eeq
where $F_{\mu\nu}=-\sqrt{2/p} f_{\mu\nu\rho}P^{\rho}$.
It differs from that of IIB matrix model by
\beq
{1\over 2}Trlog(P^2\delta_{\mu\nu})
-{1\over 2}Trlog(P^2\delta_{\mu\nu}-2iF_{\mu\nu}).
\eeq
In dimensions higher than 2, we can expand this expression into
the power series of $F_{\mu\nu}$.
The leading correction is found to be
\beq
3Tr{1\over T^2}
=3\sum_{n=1}^p(n+1)^3{1\over n(n+2)}
\sim 3N .
\label{cosmo}
\eeq
We find that the one loop correction is
of the opposite (positive) sign with the classical action.
It is unlike the $CP^1=SU(2)/U(1)$ case where the quantum corrections
vanish due to SUSY.
Nevertheless the quantum correction in $CP^2$ case is proportional to
$N$ and hence the volume of the manifold.
As a catch phrase, one might say that the quantum correction to the
cosmological constant is finite in this model.
The contribution to (\ref{cosmo}) is dominated by the states with large
Casimir eigenvalues. Such states represent nonlocal states connecting the
opposite sides of spacetime. In this sense
we believe that this quantum correction is the signature of cosmic scale
SUSY breaking and there is no contradiction to the notion of local recovery
of SUSY.

We investigate
formal semiclassical (or infrared) limit of NC gauge theory on $CP^2$.
In this case, we fix the size of spacetime to be 1.
In a semiclassical approximation, we can identify
\beqa
&&Tr \rightarrow {1\over (2\pi )^2}({p\over 2})^2\int d^4x \sqrt{g},\n
&&P_{\mu}P_{\mu}\rightarrow
-{1\over \sqrt{g}}\partial_a \sqrt{g}g^{ab}
\partial_b .
\eeqa
We may  further identify
\beq
P_{\mu}\rightarrow -iK_{\mu}^a(x)\partial_a ,
\eeq
where the Killing vectors in homogeneous spaces
are related to the inverse metric as
\beq
\sum_{\mu }K_{\mu}^aK_{\mu}^b=g^{ab} .
\eeq
Since $P_{\mu}$ satisfy the Lie algebra of $SU(3)$, we find
\beq
K_{\mu}^a\partial_a K_{\nu}^b
-K_{\nu}^a\partial_a K_{\mu}^b
=-f_{\mu\nu\rho}K_{\rho}^b .
\label{Ktrnf}
\eeq

We can also parameterize
\beq
\hat{a}_{\mu}\rightarrow K_{\mu}^a(x)b_a(x)+N_{\mu}^i(x)\phi_i(x) ,
\eeq
where $N_{\mu}^i$ are orthogonal to the Killing
vectors
\beq
\sum_{\mu}K_{\mu}^aN_{\mu}^i=0 .
\label{orth}
\eeq
In order that (\ref{orth}) is consistent with (\ref{Ktrnf}),
$N_{\mu}^i$ must satisfy
\beq
K_{\mu}^a\partial_a N_{\nu}^i
-K_{\nu}^a\partial_a N_{\mu}^i
=-f_{\mu\nu\rho}N_{\rho}^i .
\label{Ntrnf}
\eeq
We may define the inverse metric in the `transverse' space from $N_{\mu}^i$.
\beq
\sum_{\mu}N_{\mu}^iN_{\mu}^j=g^{ij} .
\eeq
From (\ref{Ktrnf}) and (\ref{Ntrnf}), we can show that $g^{ij}$ is invariant
under the  isometry
\beq
K_{\mu}^a\partial_a g^{ij}=0 .
\eeq
We can subsequently conclude that the transverse space is flat, namely
$g^{ij}=\delta^{ij}$.

Since
\beqa
&&P_{\mu}a_{\nu}-P_{\nu}a_{\mu}\n
&=&-iK_{\mu}^aK_{\nu}^b(\partial_a b_b-\partial_b b_a)
+if_{\mu\nu\rho}K_{\rho}^ab_{a}\n
&&-i(K_{\mu}^aN_{\nu}^i-K_{\nu}^aN_{\mu}^i)\partial_a \phi_i
+if_{\mu\nu\rho}N_{\rho}^i\phi_i ,
\eeqa
we find
\beq
F_{\mu\nu}=
-iK_{\mu}^aK_{\nu}^b(\partial_a b_b-\partial_b b_a)
-i(K_{\mu}^aN_{\nu}^i-K_{\nu}^aN_{\mu}^i)\partial_a \phi_i .
\eeq
The bosonic part of the action (\ref{act2})
gives the following result in the semiclassical limit.
\beqa
&&\int d^4x \sqrt{g}
\left( {1\over 4}g^{ac}g^{bd}
(\partial_a b_b-\partial_b b_a)(\partial_c b_d-\partial_d b_c)
+{1\over 2}g^{ab}\partial_a\phi_i\partial_{b}\phi_i \right. \n
&&\left. -{1\over 2}
f_{\mu\nu\rho}K^a_{\mu}K^b_{\nu}N^i_{\rho}
(\partial_a b_b-\partial_b b_a)\phi_i \right) .
\eeqa
We thus obtain gauge theory on the four dimensional
curved manifold $CP^2$.

The fermionic part is
\beq
{1\over 2}Tr\bar{\psi}\Gamma_{\mu}[A_{\mu},\psi] .
\eeq
In the semiclassical limit, we obtain
\beq
{1\over 2}\int d^4x \sqrt{g}
\left(-i\bar{\psi}\gamma^a\partial_a\psi
\right) ,
\eeq
where $\gamma^a=\Gamma_{\mu}K_{\mu}^a$ and $\gamma^i=\Gamma_{\mu}N_{\mu}^a$.
They satisfy the following commutation relations
\beq
\{\gamma^a,\gamma^b\}=g^{ab},
~~\{\gamma^i,\gamma^j\}=\delta^{ij},
~~\{\gamma^a,\gamma^i\}=0 .
\eeq
We note that the fermionic kinetic term is invariant under $SU(3)$
transformations
\beq
Tr\bar{\psi}\Gamma_{\mu}[\hat{p}_{\mu},\psi]
\rightarrow \int d^4x \sqrt{g}(-i\bar{\psi}\Gamma_{\mu}
K_{\mu}^a\partial_a \psi) .
\eeq
It is because an $SU(3)$ rotation
\beq
\delta \hat{p}_{\mu}=i \alpha f_{\mu\nu\rho}\hat{p}_{\nu}
=[\hat{p}_{\rho},\hat{p}_{\mu}]
\eeq
can be undone by the $SO(10)$ transformation of $\psi$ as
\beq
\delta\psi=-{i\over 4} \alpha f_{\mu\nu\rho}\Gamma^{\mu\nu}\psi .
\eeq
In symmetric homogeneous spaces,
we can locally choose a flat metric for $g^{ab}$
with the vanishing spin connection.
Since our fermionic action is valid locally,
it must be valid globally due to $SU(3)$ invariance.

Before concluding this section,
we make a brief comment on NC gauge theory on $CP^3$ case.
The classical action is evaluated
for the irreducible representation $(m,n)$ of $SO(5)$ as
\beq
-{\alpha^4\over 2}C_2(m,n)dim(m,n),
\eeq
where
\beqa
&&C_2(m,n)=m^2+n^2+3m+n ,\n
&&dim(m,n)={(m+n+2)(m-n+1)(3+2m)(1+2n)\over 6} .
\eeqa
It is not minimized by an $(m,m)$ type representation
but rather minimized by an
$(m,0)$ type representation for the fixed $N=dim(m,n)$.
Unfortunately the $(p/2,p/2)$ representation which is relevant to
$CP^3$ as it is explained in section 2
is metastable in our construction.
The quantum correction around the $(p/2,p/2)$
state is found to be much larger than the classical action since
\beq
6Tr{1\over T^2}
=6\sum_{m=0}^p\sum_{n=0}^m dim(m,n){1\over C_2(m,n)}
\sim {3\over 4}(log(4)-1)p^4 ,
\eeq
while $N\sim p^3/6$. So the one loop correction to the
cosmological constant is infinite in this model.

\section{Conclusions and Discussions}
\setcounter{equation}{0}
We have investigated NC gauge theories on fuzzy homogeneous
spaces $G/H$ through matrix models.
We have considered deformed IIB matrix models with finite $N$.
In our construction, the isometry of a homogeneous space,
$G$ must be a subgroup of $SO(10)$
which is the symmetry of IIB matrix model.
A local patch and the coordinates can be introduced
semiclassically when $N$ is large.
We require that a fuzzy flat hyper plane is
realized in the local patch.
Although the Hermitian matrices $A_{\mu}$ have been interpreted
as coordinates in matrix models, we have interpreted them
as Killing vectors of spacetime.

We have investigated 4 dimensional NC gauge theory on
fuzzy $CP^2$ in detail.
We have shown that NC gauge theory on $R^4$ is realized
by letting the cubic coupling vanish in the large $N$ limit.
We have therefore made the relation between matrix models and NC gauge
theory more precise.
Our construction may be useful for nonperturbative investigations
of supersymmetric NC gauge theories.
In string theory, NC gauge theory is realized by introducing
constant $B_{\mu\nu}$ field\cite{SW}\cite{Ishibashi}.
Let us consider a localized state $|p>$ on $CP^2$ as it is
discussed in section 2. In the local patch around it,
$\{t^4,t^5,t^6,t^7\}$ can be interpreted as the local coordinates.
$f_{458}$ and $f_{678}$ can be interpreted as constant
$B_{12}$ and $B_{34}$ fields since the deformation term
behaves locally like
\beqa
&&if_{458}{(2\pi )^2\over g_{NC}^2}
Tr[({2\over p}t_8) ([\hat{a}_{4},\hat{a}_{5}]+[\hat{a}_{6},\hat{a}_{7}])]\n
&\rightarrow&{i\over g_{NC}^2}\int d^4x (a_4\star a_5-a_5\star a_4
+a_6\star a_7-a_7\star a_6) .
\eeqa
This is consistent with the coupling of $B_{\mu\nu}$ field
in IIB matrix model\cite{Kitazawa}.

We hope to draw possible implications for the large
$N$ limit of IIB matrix model based on our results.
We recall that the cubic terms we have added to IIB matrix model formally
vanish in the large $N$ limit.
Nevertheless they affect the theory since different
NC gauge theories are realized based on different homogeneous
spaces $G/H$. The situation is analogous to
the magnetic systems where different polarization directions are realized
depending on the directions of a tiny external magnetic field.
We should hence clarify under what conditions unique physics
is realized in the large $N$ limit of IIB matrix model.
Another related issue is the possibility that the cubic terms
may be dynamically generated in the large $N$ limit of IIB matrix model.
We recall here that four dimensional distributions for $A_{\mu}$ are found
to be favored in the mean field approximation\cite{NS}\cite{Kyoto}.
It is interesting to study which homogeneous space
minimizes the free energy of IIB matrix model under the mean field
approximation.

\begin{center} \begin{large}
Acknowledgments
\end{large} \end{center}
I would like to thank N. Ishibashi, S. Iso,
Y. Kimura and T. Masuda for discussions.
This work is supported in part by the Grant-in-Aid for Scientific
Research from the Ministry of Education, Science and Culture of Japan.


\newpage

\begin{thebibliography}{99}
\bibitem{BFSS}
T. Banks, W. Fischler, S.H. Shenker and L. Susskind,
{\em M-theory as a matrix model: a conjecture},
Phys. Rev. {\bf D55} 5112 (1997), hep-th/9610043.
\bibitem{IKKT}N. Ishibashi, H. Kawai, Y. Kitazawa and A. Tsuchiya,
{\em A Large-N Reduced Model as Superstring},
Nucl. Phys. {\bf B498} (1997) 467, hep-th/9612115.
\bibitem{CDS} A. Connes, M. Douglas and A. Schwarz,
JHEP 9802: 003.1998, hep-th/9711162.
\bibitem{AIIKKT}
H. Aoki, N. Ishibashi, S. Iso, H. Kawai, Y. Kitazawa
and T. Tada,
{\em Non-commutative Yang-Mills in IIB matrix model},
Nucl. Phys. {\bf 565} (2000) 176,
hep-th/9908141.
\bibitem{Li}
M. Li, Nucl.Phys. {\bf B499} (1997) 149,hep-th/961222.
\bibitem{IIKK}
N. Ishibashi, S. Iso, H. Kawai and Y. Kitazawa, {\em Wilson loops in
non-commutative Yang-Mills}, Nucl. Phys. {\bf B573} (2000) 573,
hep-th/9910004.
\bibitem{Gross}
D. J. Gross, A. Hashimoto and N. Itzhaki,
{\em Observables of Non-Commutative Gauge Theories},
Adv.Theor.Math.Phys. {\bf 4} (2000) 893,hep-th/0008075.
\bibitem{MRS}
S. Minwalla, M.V. Raamsdonk and N. Seiberg,
{\em Non-commutative Perturbative Dynamics},
JHEP {\bf 0002} (2000) 020,hep-th/9912072.
\bibitem{IKK}
S. Iso, H. Kawai and Y. Kitazawa,
{\em Bi-local Fields in Non-commutative Field Theory},
Nucl.Phys. {\bf B576} (2000) 375,hep-th/0001027.
\bibitem{Suss}
L. Susskind, {\em The quantum Hall fluid and non-commutative
Chern Simons Theory},hep-th/0101029.
\bibitem{Masuda}
S. Aoyama and T. Masuda,
{\em The Fuzzy K\"{a}hler Coset Space with the Darboux Coordinates},
Phys. Lett. {\bf B 521} (2001) 376,hep-th/0109020.
\bibitem{KN}
D. Karabali and V.P. Nair,
{\em Quantum Hall Effect in Higher Dimensions},
hep-th/0203264.
\bibitem{ZH}
S.C. Zhang and J.P. Hu,
Science {\bf 294} (2001) 823;
J.P. Hu and S.C. Zhang, cond-mat/0110572.
\bibitem{Yang}
C.N. Yang, {\em SU(2) monopole harmonics},J.Math.Phys. {\bf 19} (1978)2622.
\bibitem{Ho}
P.M. Ho, S. Ramgoolam,{\em Higher Dimensional Geometries from
Matrix Brane Constructions},
Nucl. Phys. {\bf B627} (2002) 266,hep-th/0111278.
\bibitem{Fabin}
M. Fabinger,{\em Higher-Dimensional Quantum Hall Effect in
String Theory},
JHEP {\bf 0205} (2002) 637, hep-th/0201016.
\bibitem{Taylor}
J. Castelino, S. Lee and W. Taylor,
{\em Longitudinal 5-branes as 4-spheres in Matrix Theory},
Nucl.Phys.{\bf B526}(1998)334,hep-th/9712105.
\bibitem{Kimura}
Y. Kimura,
{\em Noncommutative Gauge Theory on Fuzzy Four-Sphere
and Matrix Model},
hep-th/0204256
\bibitem{IKTW}
S. Iso, Y. Kimura, K. Tanaka and K. Wakatsuki,
{\em Non-commutative Gauge Theory on Fuzzy Sphere
from Matrix Model},
Nucl Phys. {\bf B604} (2001) 121,
hep-th/0101102.
\bibitem{Nair}
V.P. Nair and S. Randjbar-Daemi,
{\em On brane solutions in M(atrix) theory},
Nucl. Phys. {\bf B533} (1998) 333,
hep-th/9802187.
\bibitem{Bonel}
G. Bonelli, {\em Matrix Strings in pp-wave babkgrounds from
deformed Super Yang-Mills Theory},
hep-th/0205213.
\bibitem{Myers}
R.C. Myers,{\em Dielectric-Branes},
JHEP 9912 (1999) 022,hep-th/9910053.
\bibitem{AMNS}
J. Ambjorn, Y. Makeenko, J. Nishimura and R. Szabo, {\em Finite N matrix
models of non-commutative gauge theory}, JHEP {\bf 9911} (1999) 029,
hep-th/9911041; {\em Non-perturbative dynamics of non-commutative gauge
theory}, Phys. Lett. {\bf B480} 399, hep-th/0002158; {\em Lattice gauge fields
and discrete non-commutative Yang-Mills theory}, JHEP {\bf 0007} (2000) 013,
hep-th/0003208.
\bibitem{SW}  N. Seiberg and E. Witten,
{\em String theory and non-commutative geometry},
JHEP {\bf 9909} (1999) 032,
hep-th/9908142.
\bibitem{Ishibashi}
N. Ishibashi,
{\em p-branes from (p-2)-branes in the Bosonic String Theory},
Nucl.Phys. {\bf B539} (1999) 107,hep-th/9804163;
{\em A Relation between Commutative and Noncommutative Descriptions
of D-branes},hep-th/9909176.
\bibitem{Kitazawa}
Y. Kitazawa,
{\em Vertex Operators in IIB Matrix Model},
JHEP {\bf 0204} (2002) 004,
hep-th/0201218.
\bibitem{NS}
J. Nishimura and F. Sugino,
{\em Dynamical Generation of Four-Dimensional Space-Time
in IIB Matrix Model},
JHEP {\bf 0205} (2002) 001,
hep-th/0111102.
\bibitem{Kyoto}
H. Kawai, S. Kawamoto, T. Kuroki, T. Matsuo, S. Shinohara,
{\em  Mean Field Approximation of IIB Matrix Model and Emergence of Four
Dimensional Space-Time},
hep-th/0204240.
\end{thebibliography}
\end{document}